\documentclass[twocolumn,amsmath,amssymb,footinbib,superscriptaddress]{revtex4-2}

\usepackage{graphicx,color}
\usepackage{dcolumn}
\usepackage{bm}
\usepackage{epstopdf}
\usepackage{amsmath}
\usepackage[colorlinks=true,citecolor=blue,urlcolor=blue,linkcolor=blue]{hyperref}
\usepackage{url}
\usepackage{xspace}

\newcommand{\mapi}{MAPbI$_{3}$\xspace}

\begin{document}

\title{Strongly Anharmonic Organic Cation Vibrations in Hybrid Lead Halide Perovskite CH$_3$NH$_3$PbI$_3$ on Electronic Ground State}

\author{Kunie Ishioka}
\email{ishioka.kunie@nims.go.jp}
\affiliation{National Institute for Materials Science, Sengen 1-2-1, Tsukuba, 305-0047 Japan}

\author{Terumasa Tadano}
\affiliation{National Institute for Materials Science, Sengen 1-2-1, Tsukuba, 305-0047 Japan}


\author{Masatoshi Yanagida}
\affiliation{National Institute for Materials Science, Namiki 1-1, Tsukuba, 305-0044 Japan}

\author{Yasuhiro Shirai}
\affiliation{National Institute for Materials Science, Namiki 1-1, Tsukuba, 305-0044 Japan}

\author{Kenjiro Miyano}
\affiliation{National Institute for Materials Science, Namiki 1-1, Tsukuba, 305-0044 Japan}

\date{\today}

\begin{abstract}

Ultrafast vibrational dynamics of inorganic-organic hybrid lead halide perovskite CH$_3$NH$_3$PbI$_3$  in the room-temperature tetragonal phase is investigated under moderate photoexcitation slightly below the band gap.  Time-resolved optical Kerr effect signal exhibits periodic modulations due to the libration and twisting of the methylammonium  molecule at 4 and 8 THz, in addition to the inorganic sublattice deformation at 1.2 THz.  These oscillations are heavily damped, and their frequencies exhibit blueshifts with increasing pump density, contrary to the expectation from conventional laser heating.  Our \textit{ab initio} lattice dynamics simulations reveal strongly anharmonic vibrational potentials for the zone-center optical phonon modes, and thereby confirm the anharmonicity to be the origin of the experimentally observed frequency blueshifts.  

\end{abstract}

\maketitle

\section{INTRODUCTION}

Organic-inorganic hybrid lead halide perovskites such as methylammonium lead iodide (CH$_3$NH$_3$PbI$_3$ or MAPbI$_3$) have attracted great attention in the past few years not only because of their excellent performances in solar cells and other opto-electronic applications but also their unusual physical properties \cite{Egger2018, Katan2018}.  
The material consists of the heavy inorganic lead halide octahedral framework (e.g. PbI$_3$) and the light organic group (e.g. CH$_3$NH$_3$ or methylammonium MA) embedded in them.  
The structural and electronic properties of the hybrid perovskites are essentially similar to those of the all-inorganic counterpart such as CsPbI$_3$.  The conduction band minimum and the valence band maximum of both materials consist primarily of Pb and halogen orbitals \cite{
Gao2016}. 
The inorganic framework is mechanically soft, and the low-frequency phonons at the Brillouin zone boundaries involving large-amplitude halogen atom motions exhibit strong anharmonicity  \cite{Brivio2015, Whalley2016, Beecher2016, Tan2017, Carignano2017, Marronnier2017, Tyson2017, Katan2018, Sharma2020, Ferreira2020}. 
This leads to mechanical instability involving structural phase transition with temperature \cite{Whitfield2016, Sharma2020}, dynamic structural fluctuation on picosecond time scale \cite{Carignano2017, Yaffe2017, Guo2017PRM}, and low thermal conductivity \cite{
Whalley2016}. 
A number of experimental and theoretical studies suggested that the inorganic lattice is deformed around a charge carrier to form a polaron, which arguably contribute to the long carrier lifetime and diffusion length of lead halide perovskites \cite{Miyata2017, Bretschneider2018, Schlipf2018, Park2018, Lan2019, Ghosh2020, Puppin2020, Liu2020PRB}.

The roles of the organic molecules in defining the material and optoelectronic properties are less straightforward and have been under extensive debate. 
Theoretical studies predicted that the larger size of the organic cations (e.g. MA$^+$) and their stronger interactions with the inorganic framework lead to the stabilization of the highly anharmonic PbI$_3$ sublattice  \cite{Quarti2014,Gao2016, Carignano2017, Tan2017} in comparison with inorganic cations (e.g. Cs$^+$).
There have been experimental and theoretical indications that the rotation of the  MA$^+$ cations within the inorganic cage is restricted in the room-temperature tetragonal phase of MAPbI$_3$, whereas the cations rotate freely in the high-temperatures cubic phase and the rotation is frozen in the low-temperature orthorhombic phase \cite{Bakulin2015, Whitfield2016, Guo2017PRM, Selig2017, Bernard2018}.  
The rotational rearrangement of the organic cations were proposed to play an important role in the charge screening and the polaron formation \cite{Leguy2015, Schlipf2018, Munson2019ACSE, Munson2019JPCC, Joshi2019, Duan2020}. 

The atomic vibrations (phonons) and their coupling with electronic states in the hybrid perovskites have been studied extensively by means of Raman scattering spectroscopy \cite{Quarti2014, Niemann2016, Leguy2016, Guo2017PRM, Perez-Osorio2018, Sharma2020, Sharma2020b},  inelastic neutron scattering \cite{Beecher2016, Druzbicki2016, Ferreira2020} and THz spectroscopy \cite{La-o-vorakiat2016, Nagai2018, Lan2019}. 
Whereas distinct phonon peaks were observed in the low-temperature phase, at room temperature the vibrational spectra typically consisted of weak broad bands, especially in the low frequency region where the vibrations of the inorganic cages and those of the rigid organic molecules appear \cite{La-o-vorakiat2016, Druzbicki2016, Nagai2018, Ferreira2020, Sharma2020}. This is because the static spectroscopic techniques detect the temporal average of the inorganic lattice the cations that fluctuate on picosecond time scale. 

An alternative approach to experimentally investigate the phonons is to excite coherent vibrations with an ultrashort optical pulse and detect them as periodic modulations of optical properties in a pump-probe scheme.  A number of time-resolved experimental studies have been performed on hybrid lead halide perovskites under electronically resonant \cite{Wang2016, Monahan2017, Ghosh2017, Fei2018, Park2018, Batignani2018, Duan2020} and non-resonant \cite{Kim2017, Batignani2018, Cheng2019, Liu2020PRB, Liu2020PRL, Guo2020} excitation conditions. 
Resonant experiments with visible pump light can provide insights on the interaction of Raman-active optical phonons at the Brillouin zone center with photoexcited carriers.  
Several studies reported coherent oscillation modes ranging from 0.6 to 5 THz for tetragonal MAPbI$_3$ at room temperature \cite{Wang2016, Monahan2017, Ghosh2017, Park2018, Fei2018, Duan2020}.
A recent two-dimensional (2D) electronic spectroscopic study \cite{Duan2020} observed an intense MA libration at 5 THz, together with a few weaker PbI$_3$ sublattice distortions below 3 THz. 
The MA libration was relatively long-lived ($>$2 ps) and its amplitude exhibited a delayed rise by 0.7 ps, which was interpreted as the coherent oscillation \emph{after} the polaron formation based on a molecular dynamics simulation.
Non-resonant experiments with infrared (IR) pump, by contrast, can offer dynamic information on the IR-active zone-center phonons in the absence of photocarriers \cite{Kim2017, Cheng2019, Guo2020, Liu2020PRL, Liu2020PRB}.  A recent THz-pump experiment \cite{Liu2020PRB} revealed short-lived ($<$1 ps) coherent oscillations of the MA restricted rotation at $\sim$5 THz for tetragonal MAPbI$_3$, together with the Pb-I stretching (LO phonon) at $\sim$4 THz, at room temperature.  

The contrast in the MA libration dynamics between the resonant and non-resonant experiments suggests qualitative difference between the electron-phonon couplings on the electronic excited and ground states.  Because the visible and IR pump lights can in principle excite different phonon modes, however, direct comparison between the two types of experiments is not necessarily straightforward.
In the present study we aim to investigate the sub-picosecond dynamics of the Raman-active phonons on the ground state of MAPbI$_3$ in the room-temperature tetragonal phase.  We perform time-resolved optical Kerr effect (TR-OKE) measurements using a near infrared pump pulse at moderate fluence, so that coherent high-frequency phonons can be monitored without creating excessive amount of photocarriers.  Periodic modulations in the TR-OKE signal by coherent  MA molecular twisting and libration are clearly detected, together with those by PbI$_6$ octahedral deformations.  
Good signal-to-noise (S/N) ratio of our signals allows quantitative analyses, which indicates the strong vibrational anharmonicities. 
Our theoretical simulations on the zone-center optical phonon modes in tetragonal MAPbI$_3$ confirm our mode assignments and the strong vibrational anharmonicity for both organic and inorganic vibrational modes.

\section{EXPERIMENTAL methods}

The sample consists of a 1-mm thick glass substrate, a 250-nm thick MAPbI$_3$ film fabricated by two-step spin coating procedure \cite{Tripathi2015, Khadka2017}, and a 100-nm thick polymethyl methacrylate (PMMA) capping layer to keep off the atmospheric humidity and oxygen.  

\begin{figure}
\includegraphics[width=0.475\textwidth]{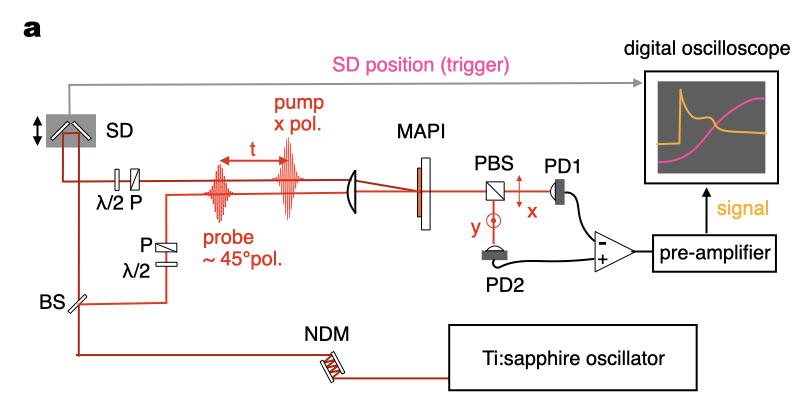}
\includegraphics[width=0.475\textwidth]{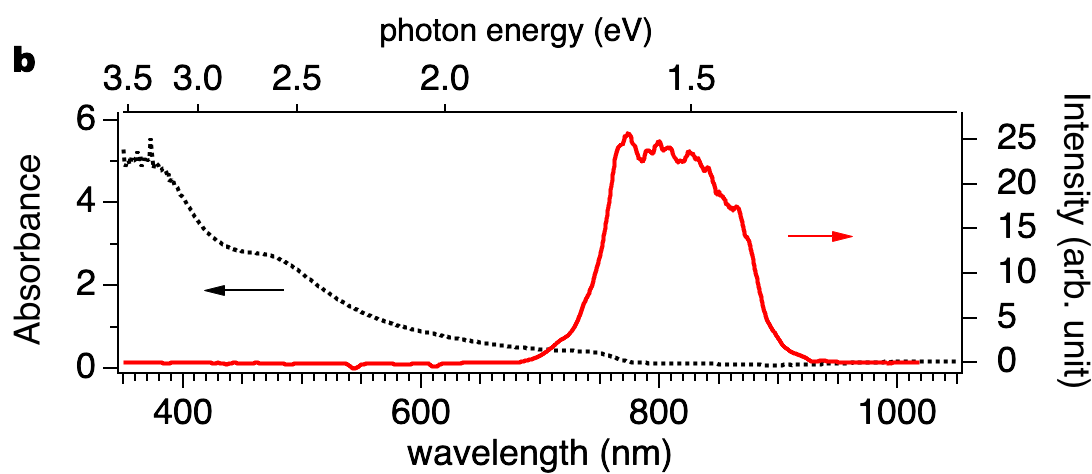}
\caption{\label{Spectra}  (a) Schematic illustration of the time-resolved optical Kerr effect (TR-OKE) measurement setup.   NDM: negative dispersion mirror pair, BS: beam splitter, SD: fast-scan delay stage, $\lambda/2$: half-wave plate, P: polarizer, MAPI: MAPbI$_3$ sample, PBS: polarizing beam splitter, PD1 and PD2: photodetectors.  (b) Spectrum of the laser pulse employed in the pump-probe measurements (red solid curve, plotted against the right axis), compared with the absorbance of MAPbI$_3$ film (black dotted curve, plotted against the left axis). 
}
\end{figure}

TR-OKE measurements are performed under ambient conditions using the setup schematically illustrated in Fig.~\ref{Spectra}a.  The output of a Ti:sapphire oscillator with 12-fs duration and 80-MHz repetition rate is used as the light source.  The laser spectrum is centered at wavelength of 810 nm (photon energy of 1.53 eV) and has a full width of 130 nm (0.24 eV), as shown in Fig.~\ref{Spectra}b.  Only a small fraction ($\sim$1/6) of the spectrum exceeds the nominal bandgap of MAPbI$_3$, 1.6 eV at room temperature.  
The maximum pump density employed in the present study,  57 $\mu$J/cm$^2$, corresponds to the incident photon flux of $F=2.3\times10^{14}$ photons/cm$^2$, and the upper limit for the photoexcited carrier density is estimated to be $3\times10^{17}$ cm$^{-3}$. 
We confirm that no irreversible photodamage is done on the sample by measuring the same spot from the lowest to the highest fluences and then again at the lowest. 
Linearly polarized pump and probe lights are focused onto the same spot of $\sim$50 $\mu$m diameter on the sample.  
Probe light is polarized at $\sim45^\circ$ from $x$ so that the horizontal ($x$) and vertical ($y$) components of the transmitted probe light is balanced on a pair of Si PIN photodetectors (S5973-01, Hamamatsu) in the absence of the pump light.
Pump-induced signal from the photodetector pair is amplified with a low noise current preamplifier (SR570, Stanford Research) and averaged with a digital oscilloscope (Waverunner HRO66Zi, LeCroy) typically over 10,000 scans, while the time delay $t$ between the pump and probe pulses is scanned at 20 Hz using a fast scan delay stage (ScanDelay50, APE). 
The averaged transient signal is finally normalized to obtain anisotropic transmissivity change, $(\Delta T_y-\Delta T_x)/T$, which we hereafter refer to as the TR-OKE signal.

\section{Experimental Results}\label{ExpResults}

\begin{figure}
\centering
\includegraphics[width=0.475\textwidth]{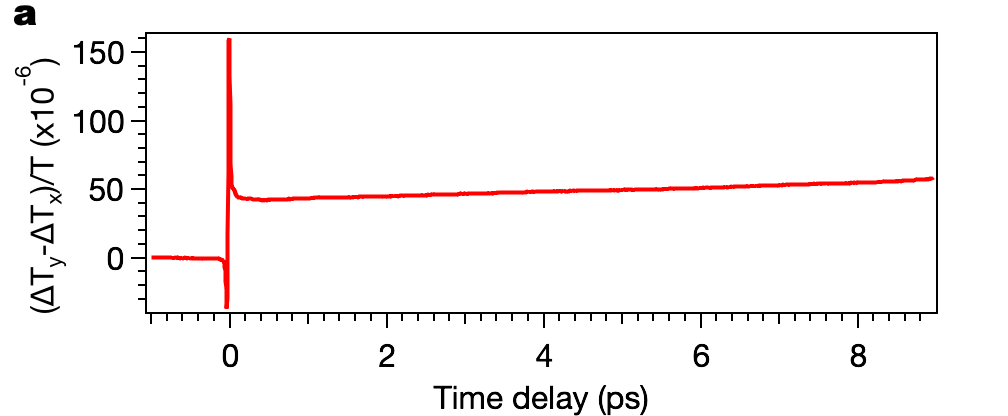}
\includegraphics[width=0.475\textwidth]{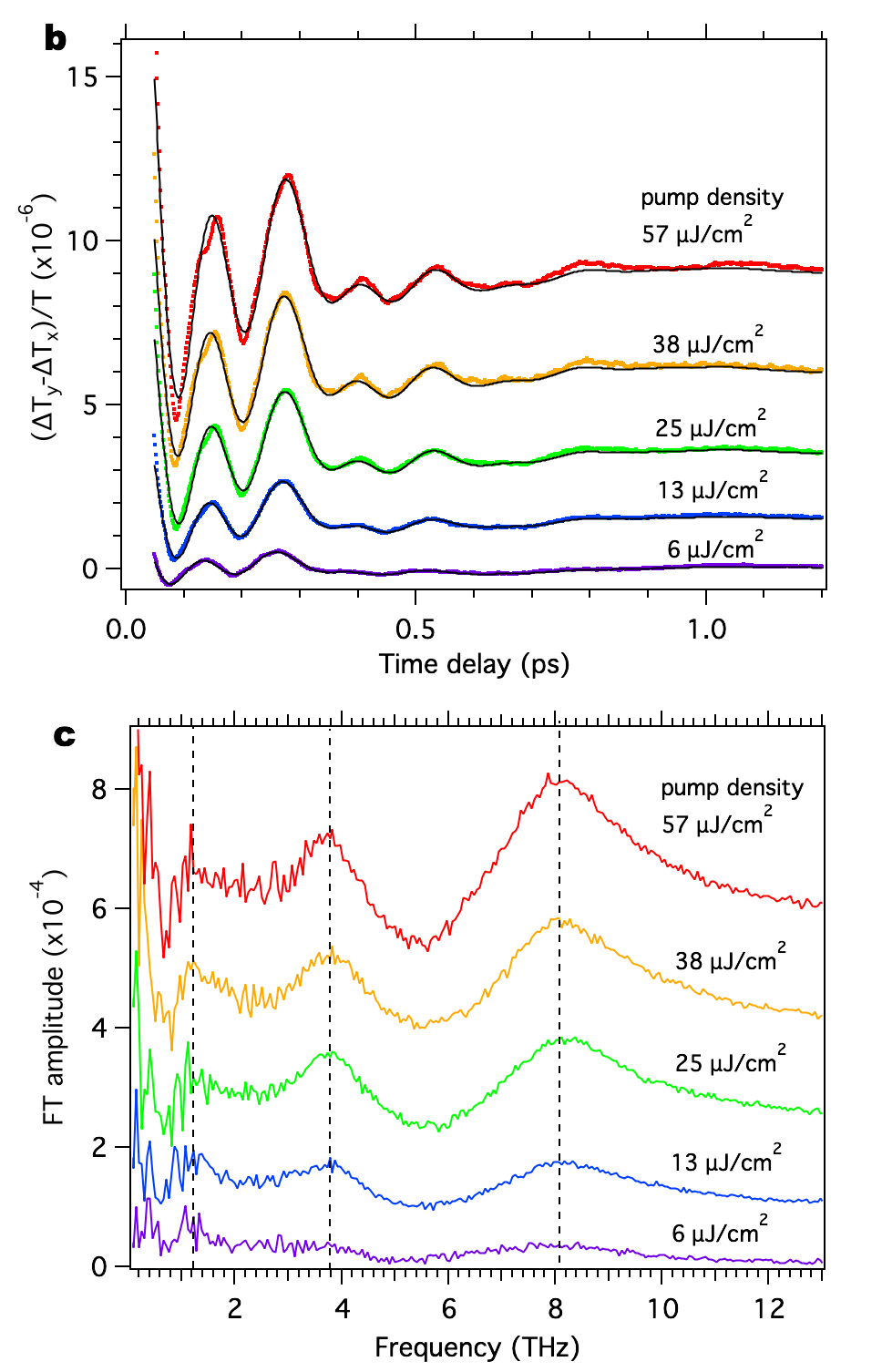}
\caption{\label{power} (a) As-measured TR-OKE signal of MAPbI$_3$ at pump density of 32 $\mu$J/cm$^2$.  (b) Oscillatory components of the TR-OKE signals pumped at different densities (colored dots) and fits to a triple damped harmonic function (black curves).  (c) Fast Fourier transform (FFT) spectra of (b). Vertical broken lines indicate the approximate positions of the FT peaks.  Traces are offset for clarity in (a) and (b).}
\end{figure}

Figure~\ref{power}a plots the as-measured TR-OKE signal of the MAPbI$_3$ film.  The signals exhibit a sharp ($<$100 fs) bipolar spike at $t$=0 followed by a positive baseline that hardly decays within the present time window of $\sim$10 ps.  The initial spike can be interpreted as the pump-induced polarizations in the PbI$_6$ octahedral framework and among the MA$^+$ cations, both of which are expected to relax on a few hundred femtosecond time scale \cite{Yaffe2017, Bakulin2015, Selig2017, Bernard2018}.  The following positive baseline indicates the long-lived photoexcited carrier population, which relaxes on nanosecond or longer time scale \cite{Ishioka2017, Kanemitsu2018, Joshi2019}.  

On top of this non-oscillatory response, a much weaker quasi-periodic modulation is also detected.  Figure~\ref{power}b shows the oscillatory component after subtraction of the non-oscillatory baseline pumped at different pump densities.  The fast Fourier transform (FFT) spectrum in Fig.~\ref{power}c shows that the coherent response is dominated by two modes at $\sim$4 and 8 THz ($\sim$130 and 260 cm$^{-1}$).  In addition, we find a weaker FT peak at 1.2 THz (40 cm$^{-1}$).

The frequency of 1.2 THz is close to that reported by previous pump-probe \cite{Kim2017, Monahan2017, Park2018} and Raman \cite{Sharma2020} studies.  This mode was attributed to I-Pb-I angular bending within the $ab$ plane based on theoretical calculations~\cite{Perez-Osorio2015, Druzbicki2016, Ponce2019, Brivio2015, Leguy2016, Park2018, Perez-Osorio2018}.
The 4-THz mode was also observed in previous Raman studies and assigned as the libration of the rigid MA$^+$ molecule based on theoretical simulations \cite{Ponce2019, Perez-Osorio2015, Park2015, Quarti2014, Brivio2015, Perez-Osorio2018, Mattoni2016, Leguy2016, Schlipf2018}.  Coherent vibrations at  a similar frequency were also reported in time-resolved spectroscopic studies on MAPbI$_3$ and MAPbBr$_3$ \cite{Batignani2018, Ghosh2017, Duan2020}.  
The 8-THz mode was assigned as the intra-molecular hindered rotation of CH$_3$ against NH$_3$ in previous Raman and theoretical studies \cite{Perez-Osorio2015, Quarti2014, Brivio2015,  Niemann2016, Leguy2016, Mattoni2016, Druzbicki2016}.  We hereafter refer to this mode as the MA twisting, though it was called differently (e.g. torsion or disrotatory motion) in the literatures.  
The MA cations are coupled with the surrounding PbI$_3$ cage via ion-ion, hydrogen bonding and ion-dipole interactions, which gives rise to the restoring force for both the twisting and librational motions.

We note that the 4- and 8-THz modes are not detected in the similar TR-OKE measurement on a single crystal MAPbI$_3$. 
This is primarily because the low damage threshold ($\sim6\mu$J/cm$^2$) of the single crystal, which makes it impossible to gain sufficient S/N ratio to observe the weak signal from the MA motions.  Below the damage threshold, the TR-OKE signal is dominated by a relatively long-living oscillation at 1.5 THz (50 cm$^{-1}$) due to the PbI$_3$ deformation.

\begin{figure}
\includegraphics[width=0.475\textwidth]{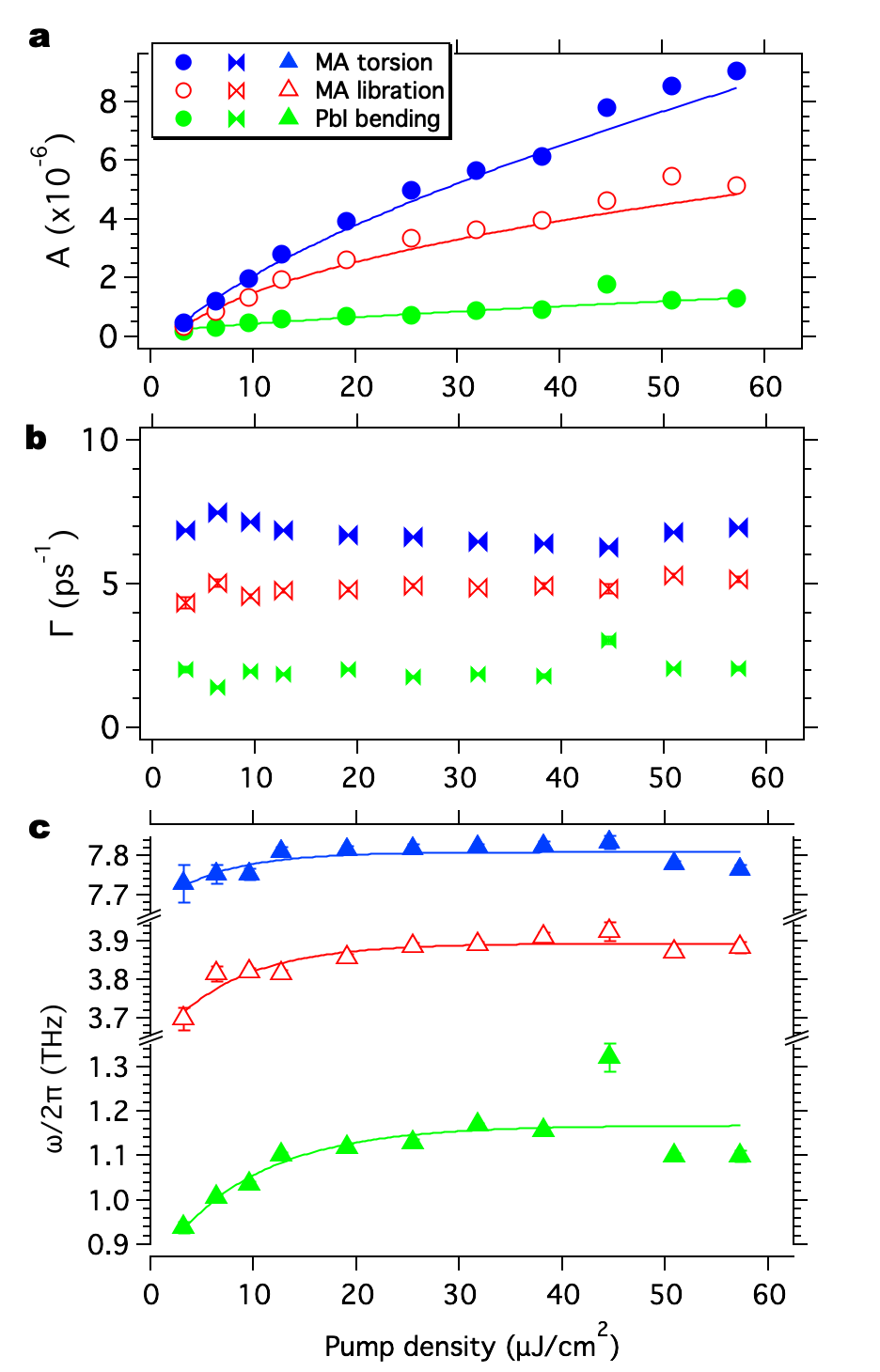}
\caption{\label{param} Initial amplitudes (a), dephasing rates (b), and frequencies (c), obtained from fitting the oscillatory TR-OKE signals to a triple damped harmonic function, as a function of pump density.  Curves are to guide the eye. 
}
\end{figure}

To extract quantitative information from the phonon dynamics of the MAPbI$_3$ thin film, we analyze the coherent oscillations by fitting to a triple damped harmonic function:
\begin{eqnarray}\label{tdh}
f_{t}(t)&=A_{b}\exp(-\Gamma_{b} t)\sin(\omega_{b} t+\phi_{b})\nonumber\\
&+A_{l}\exp(-\Gamma_{l} t)\sin(\omega_{l} t+\phi_{l})\nonumber\\
&+A_{t}\exp(-\Gamma_{t} t)\sin(\omega_{t} t+\phi_{t})
\end{eqnarray}
with the subscripts \textit{b}, \textit{l} and \textit{t} denotin the Pb-I angular bending, the MA$^+$ libration, and the MA$^+$ twisting, respectively.  
In Fig.~\ref{power}a we see that the fits (black curves) satisfactorily reproduce the experimental data (colored dots) for $t\gtrsim60$ fs.  The good fit indicates that the coherent oscillations are generated almost instantaneously, in apparent contrast to the significantly delayed rise in the MA libration amplitude due to the polaron formation upon above-bandgap photoexcitation \cite{Duan2020}.  The contrast suggests that the effect of the polaron formation on the phonon dynamics is negligible in the present study, where the photocarrier density is lower by two orders of magnitude than the previous study \cite{Duan2020}.  In other words, we can approximately regard the coherent vibrations to be taking place on the electronic ground state.

Figure~\ref{param} summarizes the fitting parameters as a function of pump density.   We see that the amplitudes of all the three vibrational modes 
in Fig.~\ref{param}a increase slightly sublinearly with pump density, indicating that the coherent vibrations are induced via one-photon process.  The dephasing rates 
in Fig.~\ref{param}b are nearly independent of pump density, supporting that the effects of the laser-induced lattice heating and of the photocarriers on the dephasings are negligible.  All the three oscillation modes are heavily damped ($\Gamma\simeq\omega$).

Most notably, the frequencies 
in Fig.~\ref{param}c exhibit blushifts with increasing pump density until they reach saturations at $\sim20 \mu$J/cm$^2$.
Such blueshift, or bond stiffening, with increasing pump density is in contrast to the typical frequency redshift, or bond softening, due to laser-induced lattice and electronic heating.  
Instead, it can in principle be explained by considering a highly anharmonic vibrational potential, whose slope becomes steeper and energy spacing becomes larger at larger atomic coordinate. 

Previous theoretical studies predicted a highly anharmonic vibrational potential for the Pb-I bending modes at the Brillouin zone boundaries for MAPbI$_3$ in the high-temperature cubic phase \cite{Beecher2016, Whalley2016, Carignano2017}.  A similarly highly anharmonic vibrational potential was also predicted for the MA spinning, i.e., rotations of the NH$_3$ and CH$_3$ groups in the same direction, in the low-temperature orthorhombic phase of MAPbI$_3$ \cite{Perez-Osorio2017}. 
Information for the zone-center phonons of MAPbI$_3$ in the room-temperature tetragonal phase, which can be directly compared with our experimental observations, has been very limited \cite{Marronnier2017, Tyson2017}, however.

\section{Theoretical simulations}

To gain deeper understanding of the zone-center optical phonons of tetragonal \mapi, we performed \textit{ab initio} lattice dynamics simulation based on density functional theory (DFT).
The DFT calculations were conducted by using the Vienna \textit{ab initio} simulation package (VASP)~\cite{kresse1996efficient}, which implemented the projector augmented wave (PAW) method~\cite{PAW1994,Kresse1999}. The experimental lattice constants of tetragonal \mapi at 300 K reported in Ref.~\cite{Lehmann2019} were employed. As for the internal coordinates, we employed the low-energy structure of Ref.~\cite{MAPIXtal} as a starting point and then relax the coordinates by DFT until the forces acting on atoms became less than 0.5 meV/\AA. 
For comparison, we also performed phonon calculations with the theoretical lattice constants obtained from DFT.  The results 
are very similar to those presented below, except that the phonon frequencies are systematically higher due to the smaller unit cell volume.
A kinetic energy cutoff of 500 eV and the 4$\times$4$\times$3 Monkhorst--Pack $\bm{k}$ points were employed. For the exchange-correlation potential we employed the PBEsol functional~\cite{Perdew2008PBEsol}, since it has better prediction accuracy of lattice constants than LDA or PBE. The harmonic phonon calculation was performed by using the finite displacement method, as implemented in the \textsc{alamode} code~\cite{Tadano2014}. 

\begin{figure}
\centering
\includegraphics[width=0.49\textwidth,clip]{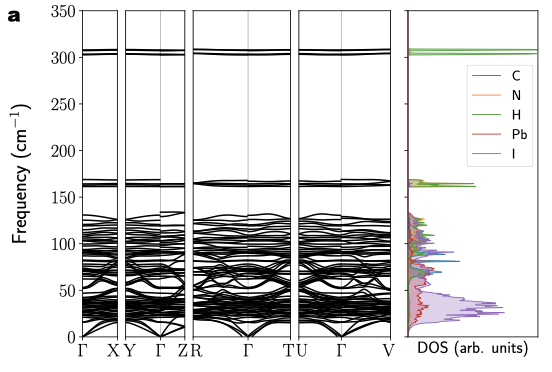}
\includegraphics[width=0.49\textwidth,clip]{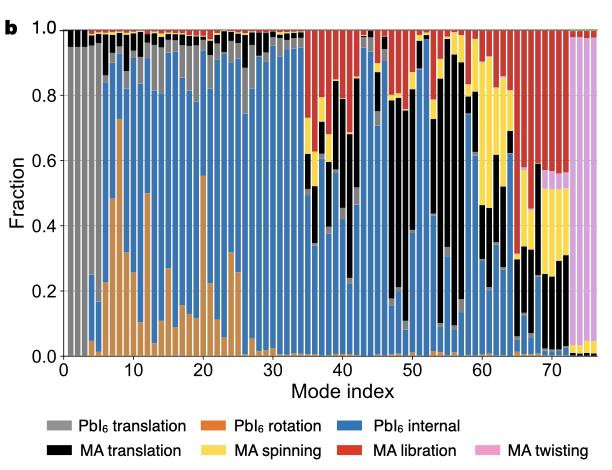}
\caption{(a) Calculated harmonic phonon dispersion curves along the high-symmetry points (left panel) and the atom-projected phonon DOS (right panel) of tetragonal \mapi in the low-frequency region. (b) Decomposition of the zone-center phonons of tetragonal MAPbI$_3$ into inorganic and organic vibrational components. The height of each colored stick represents the fractional contribution of each vibrational component.  Phonon modes are indexed in the ascending order of the harmonic frequencies.  
}
\label{fig:Dispersion} 
\end{figure}

\begin{figure*}
\centering
\includegraphics[width=0.55\textwidth]{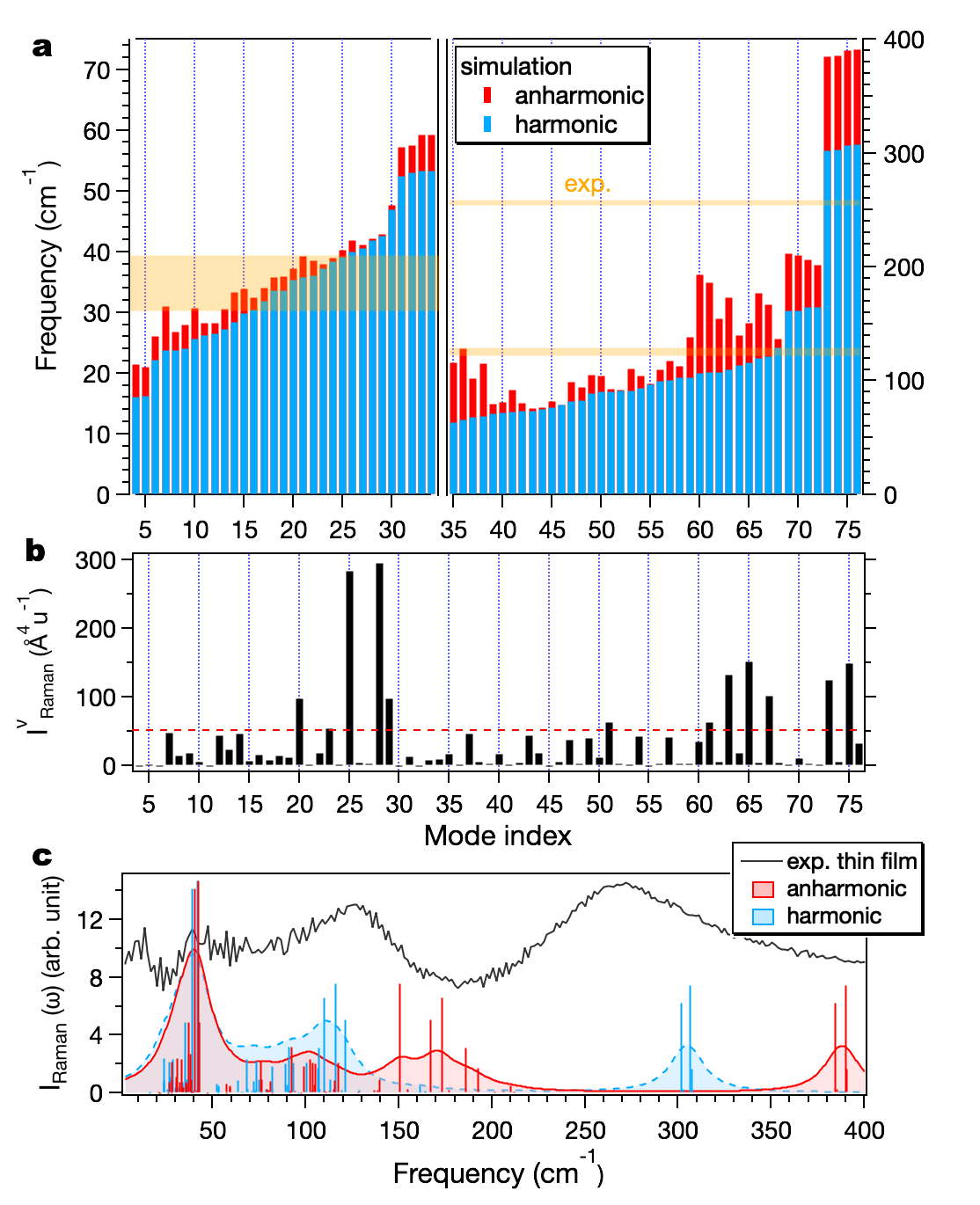}
\includegraphics[width=0.44\textwidth]{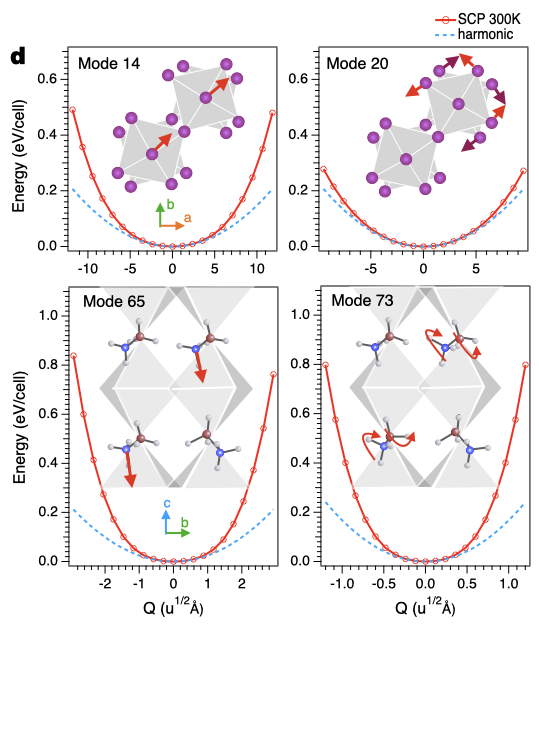}
\caption{\label{sim_raman_pes} (a) Phonon frequencies calculated with harmonic approximation (blue bars) and SCP theory assuming a temperature of 300 K (red bars).  The horizontal yellow bands indicate the frequency ranges observed experimentally.  (b) Raman polarizability $I_{\mathrm{Raman},\nu}$ averaged over the crystallographic axes.  Dashed line corresponds to $I^\nu_\text{Raman}=50 \AA^{4}u^{-1}$. 
(c) Raman intensity $I_{\mathrm{Raman},\nu}$ calculated with the linewidth $\Gamma=10$ cm$^{-1}$. 
Curves with red and blue shades denote the spectra calculated with anharmonic and harmonic frequencies.  Red and blue vertical bars indicate the anharmonic and harmonic $I_{\text{Raman},\nu}$.  Gray curve shows the experimental FT spectrum for the MAPbI$_3$ film, which is offset for clarity.
(d) Potential energy surface (PES) of selected Raman-active phonon modes. Red solid curves represent the anharmonic PES, whereas the blue dotted lines are the PES within the harmonic approximation. 
 The range of the horizontal axis is set from $-Q_{\text{max}, \nu}$ to  $Q_{\text{max}, \nu}$  for each panel.  Insets show schematic illustrations of atomic motions. 
 }
\end{figure*}

While the tetragonal phase is realized only in the temperature range of 161--330 K, the harmonic phonon theory predicts this phase to be dynamically stable even at $T = 0$ K, in accord with previous theoretical calculations.  Figure~\ref{fig:Dispersion}a shows the phonon dispersion and the atom-projected phonon density-of-states (DOS) of tetragonal \mapi calculated with a 2$\times$2$\times$2 supercell in the low frequency region below 350 cm$^{-1}$. 
In this frequency range, 76 normal modes are found at the $\Gamma$ point of the Brillouine zone.  Hereafter we index the normal modes in the ascending order of their frequencies $\omega_{\nu}$ calculated within the harmonic approximation, as shown with blue bars in Fig.~\ref{sim_raman_pes}a.  For each normal mode we decompose the displacements of the atoms into the translations, rotations, and internal vibrations of the PbI$_6$ octahedra and MA molecules, based on the approach reported in Ref.~\cite{Perez-Osorio2015}. Unlike Ref.~\cite{Perez-Osorio2015}, however, we also consider the twisting motion of MA. 
The result, shown in Fig.~\ref{fig:Dispersion}b, indicates that modes $\nu$=1--3 are pure PbI$_6$ translation (acoustic phonons).  Modes $\nu$=4--35 ($\omega_{\nu}\lesssim$ 65 cm$^{-1}$) are dominated by the PbI$_6$ rotation and stretching. Modes $\nu$=36--72 ($\omega_{\nu}$=66--165 cm$^{-1}$) are mixtures of the PbI$_6$ internal motions with the MA translation, libration and spinning. Modes $\nu$=73--76 ($\omega_{\nu}\simeq$ 300 cm$^{-1}$) are contributed almost exclusively by the MA twisting. 

In the present experiment, where the pump pulse energy is barely above the band gap of MAPbI$_3$, we expect that coherent phonons are Raman-active phonon modes, since they are excited predominantly via impulsive stimulated Raman scattering (ISRS) \cite{Dhar1994}. The detection of the coherent phonons in the transmission geometry also requires the phonon mode to have non-zero Raman polarizability \cite{Merlin1997}. 
We therefore evaluate the Raman scattering intensity of the normal modes to help assigning the experimentally observed phonon modes.  This is done by first calculating the dielectric constant tensor component $\epsilon^\infty_{jk}$ as a function of the normal coordinate displacement $Q_v$ based on density functional perturbation theory. The Raman polarizability tensor component $a_{jk}^{\nu}$ is then evaluated as~\cite{Porezag_PRB1996, Skelton_PCCP2017}: 
\begin{equation}
a_{jk}^{\nu}=\dfrac{V_0}{4\pi}\frac{\partial \epsilon^{\infty}_{jk}}{\partial Q_{\nu}}
\approx \frac{V_0}{4\pi} \frac{\epsilon_{jk}^\infty(+\Delta Q_{\nu})-\epsilon_{jk}^\infty(-\Delta Q_{\nu})}{2\Delta Q_\nu}, \label{eq:Raman_tensor}
\end{equation}
where $V_{0}$ is the unit-cell volume, and $\Delta Q_{\nu}$ is a small change in the normal coordinate displacement.  Finally, the Raman scattering intensity averaged over crystallographic axes, $I_\text{Raman}^\nu$, is calculated from $a_{jk}$~\cite{Porezag_PRB1996} to compare with our experiments on micro-crystals with random orientations.  
Figure ~\ref{sim_raman_pes}b summarizes the $I_\text{Raman}^\nu$ for  the 73 optical phonon modes.  While most of the phonon modes are Raman-active due to the low symmetry ($P1$) of the crystal structure, the Raman polarizability is particularly large ($\gtrsim 50$ \AA$^4$u$^{-1}$) for modes $\nu$=7, 14, 20, 25, 28, 29, 63, 65, 67, 73, and 75. 
The result of our harmonic phonon calculation agrees well with that of the previous theoretical prediction based on the similar harmonic model \cite{Leguy2016}.
We also calculate the Raman spectrum as a function of frequency $\omega$ by taking the sum of $I_{\text{Raman}}^\nu$ convoluted with a Lorentzian function with a finite linewidth $\Gamma$ over all the normal modes $\nu$. 
The result is shown in Fig.~\ref{sim_raman_pes}c. 

We now evaluate the anharmonicity of the vibrational potential of each phonon mode at the $\Gamma$ point.  We employ the self-consistent phonon (SCP) theory~\cite{Hooton_PhilosMag1958}, which has successfully been applied to compute anharmonic phonon frequencies of strongly anharmonic materials~\cite{Tadano_PRB2015,Tadano_PRL2018, Errea_EPJ2016, Tadano_JPSJ2018}. We first  calculate the potential energy surface (PES) of phonon mode $\nu$ as a function of atomic displacement $Q_{\nu}$. The calculated PES  is then fitted to the polynomial: 
\begin{equation}
    U(Q_{\nu})-U(0)=\dfrac{1}{2}\omega_\nu^2Q_\nu^2+\sum_{n=3}\frac{1}{n!}\Phi_\nu^{(n)}Q_\nu^n, \label{eq:taylor}
\end{equation}
with $\omega_{\nu}^{2}$ being the squared harmonic frequency.   Finally the anharmonic frequency $\Omega_{\nu}$  is obtained by solving the one-dimensional SCP equation:
\begin{equation}
\Omega_\nu^2=\omega_\nu^2+\sum_{n=1}\frac{1}{2^n}\Phi_\nu^{(2n+2)}\times \left[\dfrac{2n(\Omega_\nu, T)+1}{2\Omega_\nu}\right]^{n}. 
\label{eq:scp}
\end{equation}
Here $n(\Omega_\nu, T)$ denotes the Bose-Einstein distribution function at phonon frequency $\Omega_\nu$ and temperature $T$, which is related with the normal coordinate amplitude $Q_\nu$ as:
\begin{equation}\label{eq:Qv}
\langle Q_\nu^2(T)\rangle=\dfrac{\hbar}{2\Omega_\nu}\left[2n(\Omega_\nu, T)+1\right]. 
\end{equation}
In the present study we  consider anharmonic terms up to the sixth order ($n=4, 6$), which  is sufficient to reach convergence of $\Omega_{\nu}$. 
Figure~\ref{sim_raman_pes}d compares the calculated anharmonic and harmonic PES for a few representative modes with relatively large Raman intensities.  Here the horizontal scale is set from  $-Q_{\text{max}, \nu}$ to $Q_{\text{max}, \nu}$, where:
\begin{equation}\label{eq:Qmax}
Q_{\text{max}, \nu}\equiv4\sqrt{\langle Q_\nu(T=300 K)^2\rangle},
\end{equation}
is four times the mean square of the displacement $Q_{\nu}$ at room temperature calculated with the harmonic frequency $\omega_{\nu}$.
%
We find that the anharmonic PES have higher energy than the harmonic one at large $Q$ for all the 73 phonon modes investigated.  This leads to the higher anharmonic frequencies (red bars in Fig.~\ref{sim_raman_pes}a) than the harmonic ones (blue bars in the figure). 
The calculated Raman spectrum is also modified by taking into account the anharmonic effect, as shown in Fig.~\ref{sim_raman_pes}c.

\section{DISCUSSION}

Comparison between the experimental results with the theoretical simulations enables us to specify the atomic motions corresponding to the coherent phonons and discuss their vibrational anharmonicities. 
We associate the experimental 8-THz mode to the simulated modes $\nu=$73 and 75, which have higher frequencies than the experiment but relatively large Raman intensities ($\gtrsim 50$ \AA$^4$u$^{-1}$).  Their atomic motions involve almost exclusively MA twisting.  The overestimation in the simulated twisting frequencies may be attributed to the approximations we have employed in our calculations, such as neglecting the intermode coupling.
The experimental 4-THz mode can be related to the modes $\nu=$63, 65, and 67 in the simulation, whose anharmonic and harmonic frequencies lie around 120--130 cm$^{-1}$ (3.6--3.9 THz). Modes $\nu=$65 and 67 are contributed mostly by the MA libration; mode $\nu=$63 is a mixture of MA libration, spinning, and PbI$_6$ internal vibrations.
The experimental 1.2-THz mode can be related to the modes $\nu=$14 and 20, 25, 28, and in the simulation, whose frequencies lie between 30 and 50 cm$^{-1}$ (0.9 and 1.5 THz).  These modes are mixtures of the PbI$_6$ internal vibrations (angular bendings) and rotations.  Our simulations thus confirm the tentative vibrational assignments we employed in Section~\ref{ExpResults}.

The calculated PES for the phonon modes mentioned above 
reveal strong vibrational anharmonicity except for $\nu$=28 and 29.  The anharmonicity of each phonon mode can also be inferred from the difference between the harmonic and the SCP frequencies shown in Fig.~\ref{sim_raman_pes}a. Our theoretical results support that the strong vibrational anharmonicity is the origin of the experimental frequency blueshift at larger vibrational amplitude, as well as of the extremely fast dephasing of the zone-center phonons.  In our simulation, the spinning, libration, and twisting of the organic molecules exhibit particularly strong anharmonicity in this order.  The strong anharmonicity for the spinning is in line with the previous theoretical study on orthorhombic MAPbI$_3$ \cite{Perez-Osorio2017}.  
For the inorganic vibrations, modes $\nu=$14, 20, and 25, which show stronger anharmonicity than $\nu=$28 and 29, are contributed more by the the PbI$_6$ rotation.
This suggests that the rotational motions are responsible for the anharmonicity of the inorganic sublattice. 

In the present simulation we have neglected the effect of the photocarriers altogether.  We consider this is a reasonable approximation, because in our experiments we excite carriers at moderate density of $\sim10^{17}$ cm$^{-1}$.  Most of the previous resonant experimental studies, by contrast, created photocarriers at higher density by one to two orders of magnitude \cite{Monahan2017, Ghosh2017, Duan2020}.  The high-density photocarriers can lead to the efficient polaron formation, or the localization of an electron on a Pb atom, on sub-picosecond time scale \cite{Bretschneider2018, Duan2020}.  This sudden change in the energy landscape could lead to the efficient excitation of coherent phonons via displacive mechanism on the electronic excited state, as theoretically predicted to be particularly effective for the skeletal Pb-I bending mode at 20 cm$^{-1}$ (0.65 THz) \cite{Park2018}.  This is not the case in the present study, in which we expect the stimulated Raman process to dominate the coherent phonon generation.

\section{conclusion}

We have experimentally and theoretically investigated the zone-center optical phonons on the ground electronic state of tetragonal MAPbI$_3$ at room temperature.   Experimental TR-OKE signals were dominated by coherent vibrations of the libration and twisting of the organic cations, which exhibited no indication of delayed rise in their amplitudes in contrast to a previous resonant experiment.  Instead, their frequencies blueshifted with increasing pump density.  Our DFT calculations on the PES have confirmed that the blueshifts are experimental manifestation of the strong vibrational anharmonicity.  We thus successfully revealed the intrinsic ultrafast vibrational dynamics of MAPb$_3$ in the absence of photocarriers, which was found to be in stark contrast to that in the presence of photocarriers.

\begin{acknowledgments}
The authors thank Dhruba B. Khadka for discussion and John H. Dunlap for single crystal sample preparation. 
\end{acknowledgments}

\bibliographystyle{apsrev4-1}
\bibliography{MAPI_PR}

\end{document}